\documentclass[aps,twocolumn,prl,showpacs,10pt]{revtex4-1} 
\usepackage{setspace}
\usepackage{graphicx}
\usepackage{subfigure}
\usepackage{epsfig}
\usepackage{amsmath}
\usepackage{epsfig}

\begin{document}

\title{Quantum Quenches in the Thermodynamic Limit}

\author{M. Rigol}
\affiliation{Department of Physics, The Pennsylvania State University,
University Park, Pennsylvania 16802, USA}

\begin{abstract}
We introduce a linked-cluster based computational approach that allows one to study quantum quenches 
in lattice systems in the thermodynamic limit. This approach is used to study quenches in 
one-dimensional lattices. We provide evidence that, in the thermodynamic limit, thermalization 
occurs in the nonintegrable regime but fails at integrability. A phase transitionlike behavior 
separates the two regimes.
\end{abstract}


\maketitle

Studies of the quantum dynamics of isolated systems are providing fundamental 
insights into how statistical mechanics emerges under unitary time evolution 
\cite{rigol_dunjko_08_34,cazalilla_rigol_10_46,dziarmaga_10,polkovnikov_sengupta_review_11}. 
Thermalization seems ubiquitous, but experiments with ultracold gases have shown that it need 
not always occur \cite{kinoshita_wenger_06,gring_kuhnert_12}, particularly near an integrable 
point \cite{rigol_dunjko_07_27,rigol_09_39,rigol_09_43}. A major goal in those studies
is to understand how to describe observables after relaxation. 
If the initial state is characterized by a density matrix $\hat{\rho}^I$ and the 
dynamics is driven by a time-independent Hamiltonian $\hat{H}$, then the time evolution of an 
observable $\hat{O}$ is given by $O(\tau)=\text{Tr}[\hat{\rho}(\tau)\hat{O}]$, where 
$\hat{\rho}(\tau)=\exp[-i\hat{H}\tau/\hbar]\hat{\rho}^I\exp[i\hat{H}\tau/\hbar]$.
A fascinating consequence of unitary dynamics is revealed by calculating the infinite-time 
average of $O(\tau)$, $\overline{O(\tau)}=\text{lim}_{\tau'\rightarrow\infty}1/
\tau'\int_0^{\tau'} d\tau\,O(\tau)=\text{Tr}[\overline{\hat{\rho}(\tau)}\hat{O}]$, 
where $\overline{\hat{\rho}(\tau)}=\text{lim}_{\tau'\rightarrow\infty}1/
\tau'\int_0^{\tau'} d\tau\,\hat{\rho}(\tau)$. If the eigenvalues $\varepsilon_\alpha$ of 
$\hat{H}$ are nondegenerate ($\hat{H}|\alpha\rangle=\varepsilon_\alpha|\alpha\rangle$, 
$|\alpha\rangle$ being the energy eigenstates) one realizes that 
$\overline{\hat{\rho}(\tau)}=\sum_\alpha W_\alpha |\alpha\rangle\langle\alpha|$, 
where $W_\alpha$ are the diagonal matrix elements of $\hat{\rho}^I$ in the energy 
eigenbasis. This means that $\overline{O(\tau)}=\sum_\alpha W_\alpha\, O_{\alpha}$, 
with $O_{\alpha}=\langle\alpha|\hat{O}|\alpha\rangle$, depends on the initial state 
through the values of the exponentially large number of parameters $W_\alpha$. 
This is to be contrasted to traditional statistical mechanics ensembles, which are 
constructed using a few additive conserved quantities of the dynamical 
system, and are expected to describe observables after relaxation.

The potential disagreement between the outcomes of unitary dynamics and statistical
mechanics is experimentally relevant \cite{kinoshita_wenger_06,gring_kuhnert_12},
particularly in the context of quantum quenches 
\cite{rigol_dunjko_08_34,cazalilla_rigol_10_46,dziarmaga_10,polkovnikov_sengupta_review_11,trotzky_chen_12}. 
In a quantum quench, the initial (pure or mixed) state with $\hat{\rho}^I$ is selected to be stationary 
under a Hamiltonian $\hat{H}^I$, and at time $t=0$ the Hamiltonian is suddenly changed to 
$\hat{H}\neq\hat{H}^I$. Computational studies have shown that, after a quench, 
observables can relax to their infinite-time averages in realistic time scales 
\cite{rigol_dunjko_08_34,rigol_09_39,rigol_09_43}. Furthermore, it has been proved that such relaxation 
occurs under very general conditions \cite{reimann_08,linden_popescu_09}. The ensemble defined 
by $\overline{\hat{\rho}(\tau)}$ is known as the diagonal ensemble (DE) \cite{rigol_dunjko_08_34}. 
[In what follows, we use the notation $\hat{\rho}_{\text{DE}}\equiv\overline{\hat{\rho}(\tau)}$ and 
$O_{\text{DE}}\equiv \overline{O(\tau)}$.] Strikingly, for few-body observables in nonintegrable 
systems, it has been found that the predictions of the DE and of statistical mechanics ensembles 
are very close to each other, with differences that decrease with increasing system size 
\cite{rigol_dunjko_08_34,rigol_09_39,rigol_09_43}. This indicates that relaxation to the 
statistical mechanics predictions, namely, thermalization, can occur even under unitary 
dynamics, and has been understood to be the result of eigenstate 
thermalization \cite{deutsch_91,srednicki_94,rigol_dunjko_08_34}.

A fundamental limitation hampering progress in this field is the 
lack of general approaches to studying quenches in large system sizes. 
Computational studies of generic (nonintegrable) models are limited to small systems, 
for which arbitrarily long times can be calculated \cite{rigol_09_39,rigol_09_43}, or short times, 
for which large or infinite system sizes can be solved
\cite{kollath_lauchli_07,manmana_wessel_07,eckstein_kollar_09,banuls_cirac_11,essler_kehrein_14}.
Consequently, what happens in the thermodynamic limit after long times has been 
inaccessible to theoretical studies.
Here, we introduce a linked-cluster based expansion for lattice models 
that overcomes that limitation enabling calculations in the DE in the thermodynamic limit. 
In linked-cluster expansions, the expectation value of an extensive observable 
$\hat{\mathcal{O}}$ per lattice site ($\mathcal{O}$) in the thermodynamic limit is computed 
as the sum over contributions of all clusters that can be embedded on the lattice 
\cite{oitmaa_hamer_book_06,supmat}. 
At the core of these expansions lies the calculation of $\mathcal{O}$ in each cluster 
$c$, with density matrix $\hat{\rho}_c$, 
$\mathcal{O}(c)={\textrm{Tr} [\hat{\mathcal{O}}\,\hat{\rho}_c]}/{\textrm{Tr} [\hat{\rho}_c]}$. 

In thermal equilibrium, the ensemble used 
when calculating $\mathcal{O}(c)$ is the grand-canonical ensemble (GE). 
Hence, $\hat{\rho}_c\equiv\hat{\rho}^\text{GE}_c=e^{-(\hat{H}_c-\mu \hat{N}_c)/k_B T}/
\textrm{Tr} [e^{-(\hat{H}_c-\mu \hat{N}_c)/k_B T}]$, where $\hat{H}_c$ and $\hat{N}_c$ are the 
Hamiltonian and total number of particle operators in cluster $c$, respectively, $\mu$ is the 
chemical potential, $k_B$ is the Boltzmann constant (set to unity in what follows), and $T$ is 
the temperature. It is common to expand $e^{-(\hat{H}_c-\mu \hat{N}_c)/T}$ in powers of $1/T$, 
which leads to the so-called high-temperature expansions (HTEs) \cite{oitmaa_hamer_book_06}. 
However, one can instead calculate $\mathcal{O}(c)$ using full exact diagonalization 
\cite{rigol_bryant_06_25}. The resulting expansions, called numerical linked-cluster 
expansions (NLCEs), have been shown to converge to lower temperatures than HTEs 
(up to the ground state in some cases) 
in various spin \cite{rigol_bryant_06_25,rigol_bryant_07_30,tang_khatami_13} and 
itinerant models \cite{rigol_bryant_07_31,tang_paiva_12_77}. 

In this work, we introduce NLCEs for the diagonal ensemble. We assume that 
the system is initially in thermal equilibrium in contact 
with a reservoir, so that the density matrix of any cluster $c$ can be written as 
$\hat{\rho}^I_c=(\sum_a e^{-(E_a^c-\mu_I N_a^c)/{T_I}} |a_c\rangle \langle a_c|)/Z^I_c$, 
where $|a_c\rangle$ ($E_a^c$) are the eigenstates (eigenvalues) of the initial Hamiltonian 
$\hat{H}_c^I$ in $c$, and $N_a^c$ is the number of particles in $|a_c\rangle$ 
($\hat{N}_c|a_c\rangle=N_a^c|a_c\rangle$, when $[\hat{N}_c,\hat{H}_c^I]=0$). 
$\mu_I$, $T_I$, and $Z^I_c=\sum_a e^{-(E^c_a-\mu^I N_a^c)/{T_I}}$ are the initial chemical 
potential, temperature, and partition function, respectively. At the time of the quench 
$\hat{H}^I_c\rightarrow\hat{H}_c$, the system is detached from the reservoir so 
that the dynamics is unitary. Writing the eigenstates of $\hat{H}^I_c$ in terms of 
the eigenstates of $\hat{H}_c$, one can define the DE in each cluster. Its density matrix reads 
$\hat{\rho}^\text{DE}_c=\sum_\alpha W^c_\alpha|\alpha_c\rangle \langle\alpha_c|$, where 
$W_\alpha^c =(\sum_a e^{-(E^c_a-\mu_I N^c_a)/{T_I}}|\langle\alpha_c|a_c\rangle|^2)/Z^I_c$, 
and $|\alpha_c\rangle$ are the eigenstates of $\hat{H}_c$ 
($\hat{H}_c|\alpha_c\rangle=\varepsilon^c_\alpha|\alpha_c\rangle$). Taking $\hat{\rho}_c$ 
in the calculation of $\mathcal{O}(c)$ to be $\hat{\rho}^\text{DE}_c$, NLCEs can be used
to compute observables in the DE.

\begin{figure}[!t]
    \includegraphics[width=0.475\textwidth]{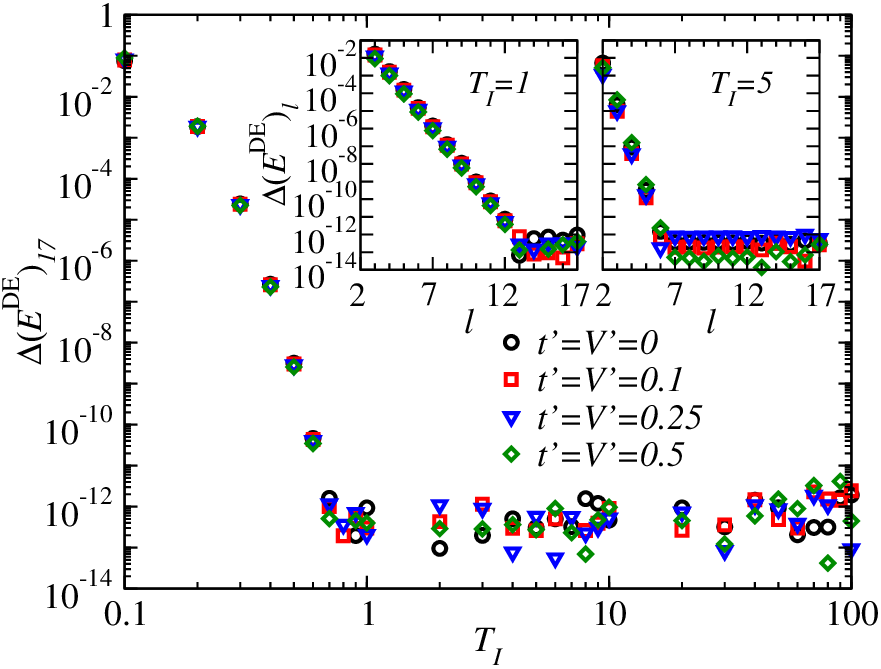}
\vspace{-0.1cm}
\caption{(color online). Relative difference $\Delta(E)_{17}$ between the last 2 orders 
in the NLCE calculation of $E^\text{DE}$ vs $T_I$ for different values of $t'=V'$ in the 
final Hamiltonian. For $T_I\gtrsim0.7$, $\Delta (E^\text{DE})_{17}$ is zero within 
machine precision. (Insets) $\Delta(E)_l$ vs $l$ for two values of $T_I$ and the 
same quenches as in the main panel. The plots show that $E^\text{DE}_{l}$ approaches 
$E^\text{DE}_{18}$ exponentially fast with $l$. 
}\label{fig:energyconv}
\end{figure}

We use these NLCEs to study quenches of hard-core bosons
in one-dimensional lattices, with nearest (next-nearest) neighbor hopping $t$ ($t'$) 
and repulsive interaction $V$ ($V'$) \cite{supmat}. This model is integrable if $t'=V'=0$ 
and nonintegrable otherwise \cite{cazalilla_citro_review_11_64}. It has been previously 
considered in quenches in finite lattices~\cite{rigol_09_39}, and in studies of the 
integrability to quantum chaos transition~\cite{santos_rigol_10_45}. After the quench, 
we take $V=t=1$ ($t=1$ sets our unit of energy), while $t'=V'$ are tuned between 
0 and 1. Unless otherwise specified, the initial state is taken to be in thermal equilibrium 
with temperature $T_I$ for $t_I=0.5$, $V_I=1.5$, and $t'_I=V'_I=0$. We restrict our analysis to 
half-filling (the average number of particles is one half the number of lattice sites). 
Given the particle-hole symmetry of our model, this is enforced by taking $\mu_I=0$.
The NLCE is implemented using maximally connected clusters, i.e., for any given 
number of lattice sites $l$, only the cluster with $l$ contiguous sites 
is used \cite{supmat}. 

\begin{figure}[!t]
    \includegraphics[width=0.465\textwidth]{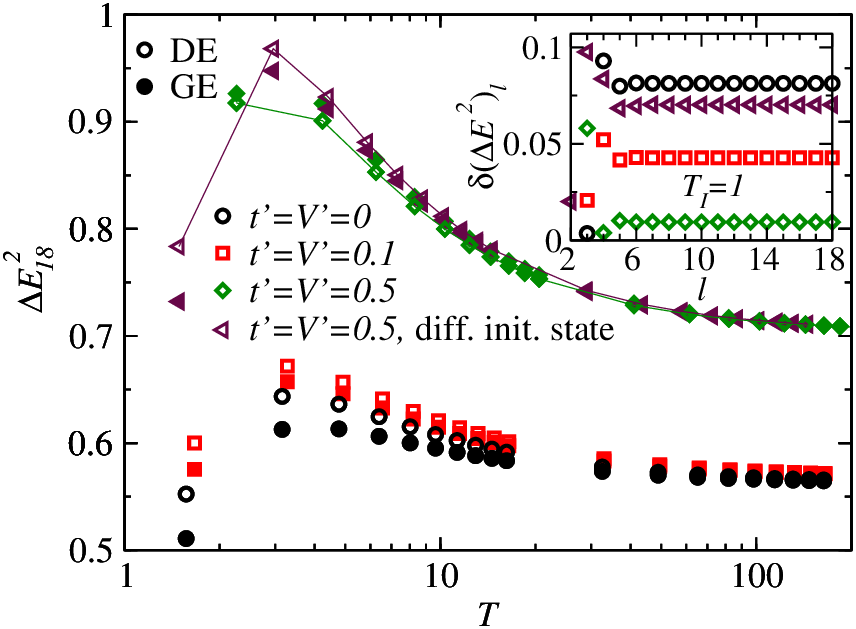}
\vspace{-0.1cm}
\caption{Last order ($l=18$) of the NLCE of $\Delta E^2$ in the DE (open symbols) 
and the GE (filled symbols) vs $T$, for quenches with 
$T_I\geq 1$. The quenches with a different initial state ($t'=V'=0.5$) have 
$t_I=0.5$, $V_I=1.5$, $t'_I=V'_I=0.5$, while $t=V=1$ as in the 
other quenches. Lines joining the data points for the quenches with $t'=V'=0.5$ 
are meant to guide the eye and show that $\Delta E^2$ in the DE depends on
the initial state. The inset shows $\delta(\Delta E^2)_l$ vs $l$ for 
$T_I=1$ for the same quenches as in the main panel.
}\label{fig:fluctuations}
\end{figure}

\begin{figure*}[!t]
    \includegraphics[width=0.85\textwidth]{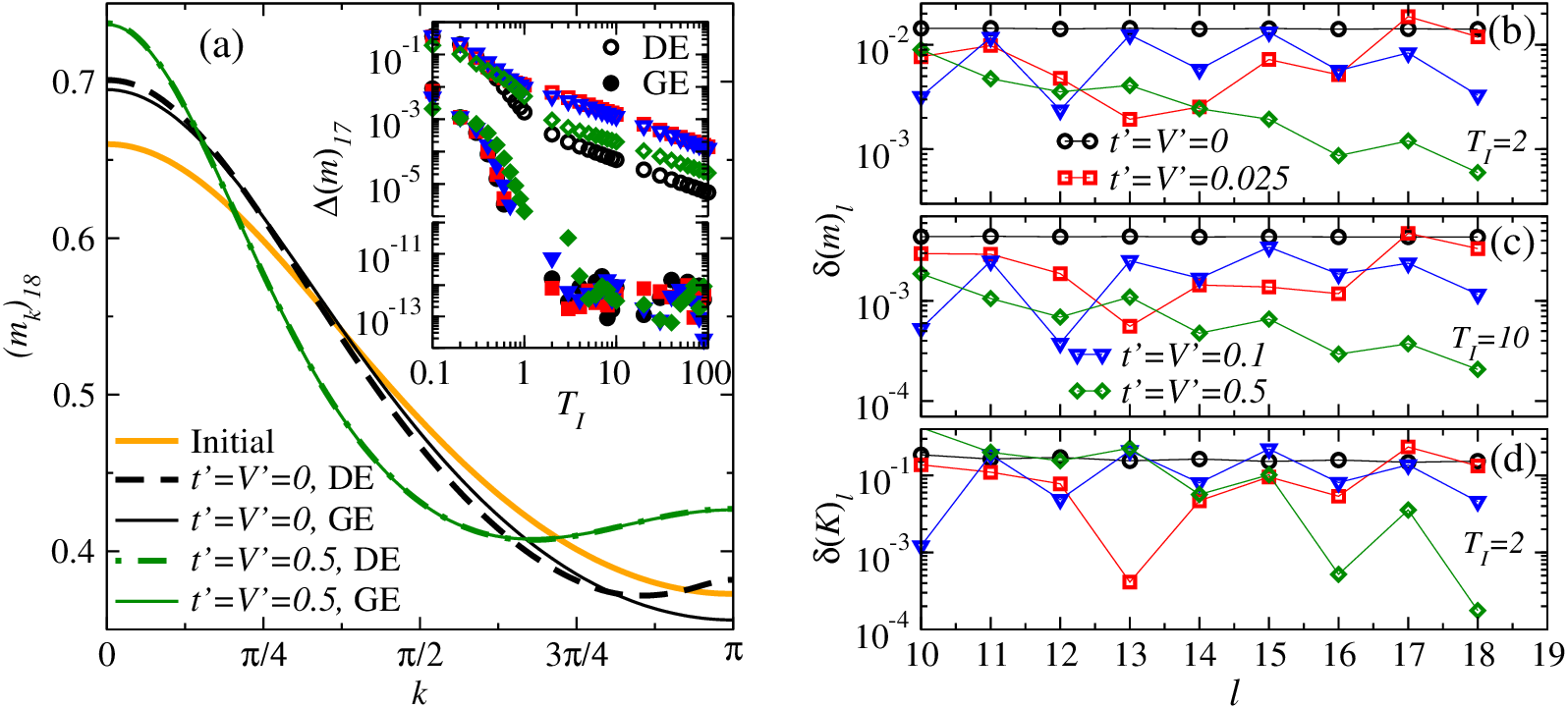}
\vspace{-0.1cm}
\caption{{\bf Momentum distribution and kinetic energy $K$ after a quench.} 
(a), Last order ($l=18$) of the NLCE for the momentum distribution in the initial state 
($T_I=2$), and in the DE and the GE after quenches with $t'=V'=0$ and 0.5. (b)--(d), 
Relative differences between $(m_k)^\text{DE}_l$ and $(m_k)^\text{GE}_{18}$ [(b),(c)] 
and between $K^\text{DE}_l$ and $K^\text{GE}_{18}$ (d) vs $l$ in four sets of 
quenches with $T_I=2$ [(b),(d)] and $T_I=10$ (c). These results provide strong 
evidence that thermalization occurs in nonintegrable systems but fails at integrability.
Inset in (a), open (filled) symbols report the relative difference between $(m_k)^\text{DE}_l$ 
[$(m_k)^\text{GE}_l$] in the last 2 orders of the NLCE vs $T_I$, for the same 
quenches as in (b)--(d). For ease of display, there is a four decade gap in the $y$ axis 
of the inset.
}\label{fig:moment}
\end{figure*}

After a quench, it is important to accurately determine the mean energy per site 
in the DE ($E^\text{DE}$). It defines, along with the mean number of particles per site 
(fixed here to be 1/2), the thermal ensemble used to determine whether observables thermalize. 
Since observables within NLCEs are computed using a finite number of clusters, we denote 
the result obtained when adding the contribution of all clusters with up to $l$ sites 
as $\mathcal{O}^\text{ens}_l$ (the superscript ``ens'' is used for DE or GE). 
To assess how close $\mathcal{O}^\text{ens}_l$ is to the thermodynamic limit result, 
we compute the difference between $\mathcal{O}^\text{ens}_l$ and the result for the 
highest order available ($l=18$ in our calculations)
\begin{equation}
 \Delta (\mathcal{O}^\text{ens})_l=\frac{|\mathcal{O}^\text{ens}_l-\mathcal{O}^\text{ens}_{18}|}
 {|\mathcal{O}^\text{ens}_{18}|}.\label{eq:D}
\end{equation}
When $\Delta (\mathcal{O}^\text{ens})_l$ becomes independent of $l$, and zero within machine 
precision, we expect that $\mathcal{O}^\text{ens}_l$ has converged to the thermodynamic limit 
result.

The accuracy of our calculation for $E^\text{DE}$ can be inferred from 
Fig.~\ref{fig:energyconv}, where we plot $\Delta (E^\text{DE})_{17}$
vs $T_I$ for several quenches. For $T_I\gtrsim0.7$, $E^\text{DE}_{17}=E^\text{DE}_{18}$ 
within machine precision. The insets in Fig.~\ref{fig:energyconv} depict 
$\Delta (E^\text{DE})_{l}$ vs $l$ for $T_I=1$ and 2. These plots show that 
(i) $E^\text{DE}_l$ approaches $E^\text{DE}_{18}$ exponentially fast with $l$, 
and (ii) with increasing $T_I$, fewer orders are required for $E^\text{DE}_l$ 
to converge to an $l$-independent result (expected to be $E^\text{DE}$ in the 
thermodynamic limit) within machine precision. The exponential convergence of 
$E^\text{DE}_l$ with $l$ shows that NLCEs are fundamentally different
from exact diagonalization. In the latter, results usually converge as a 
power law in system size~\cite{supmat}.

Once $E^\text{DE}$ is known, one can define an effective temperature after the quench 
($T$) as that of a grand-canonical ensemble such that $E^\text{GE}=E^\text{DE}$. (Here, 
all effective temperatures are computed, enforcing that the relative energy difference 
between $E^\text{DE}_{18}$ and $E^\text{GE}_{18}$ is smaller than $10^{-11}$.) 
A question that arises is whether one can make simple measurements in a system after 
a quench that will distinguish it from one in thermal equilibrium. The dispersion of 
the energy per site $\Delta E^2=(\langle\hat{H}^2\rangle-\langle\hat{H}\rangle^2)/L$ are a good 
candidate (see Ref.~\cite{supmat} for another one). In thermal equilibrium they depend 
on the ensemble used to compute them. $\Delta E^2=0$ in the microcanonical ensemble while
$\Delta E^2\geq0$ in the canonical ensemble and the GE. $\Delta E^2$ is also of 
interest because, in the latter two ensembles, the specific heat 
$C_v=(1/L)\partial\langle\hat{H}\rangle/\partial T \propto\Delta E^2$.

In what follows, in order to quantify how order by order the DE prediction for 
an observable compares to the GE result in the last order, we define the relative 
difference
\begin{equation}
 \delta (\mathcal{O})_l=\frac{|\mathcal{O}^\text{DE}_l-\mathcal{O}^\text{GE}_{18}|}
 {|\mathcal{O}^\text{GE}_{18}|}.\label{eq:d}
\end{equation}
We make sure that, for all results reported for $\delta (\mathcal{O})_l$, the analysis of 
$\Delta(\mathcal{O}^\text{GE})_l$ suggests that $\mathcal{O}^\text{GE}_{18}$
has converged to the thermodynamic limit result.

In the main panel in Fig.~\ref{fig:fluctuations}, we plot $\Delta E^2$ 
in the DE (empty symbols) and in the GE (filled symbols) vs $T$ for quenches
with $T_I\geq 1$. These results, particularly the ones at the lowest temperatures, 
make it apparent that $\Delta E^2$ is different in the DE and the GE.
Moreover, as shown in Fig.~\ref{fig:fluctuations} for quenches with different 
initial states but the same final Hamiltonian ($t'=V'=0.5$), $\Delta E^2$ in the 
DE depends on the initial state. The relative differences $\delta(\Delta E^2)_l$, 
between $\Delta E^2$ in the DE for order $l$ and in the last order 
in the GE are plotted in the insets in Fig.~\ref{fig:fluctuations}(a) vs $l$. 
They show that the nonzero differences seen in the main panels between 
$\Delta E^2$ in the DE and the GE are fully converged and are thus expected to be 
the ones in the thermodynamic limit. The fact that $\Delta E^2$ in the DE and the GE agree 
with each other as $T_I\rightarrow\infty$ (main panel in Fig.~\ref{fig:fluctuations}) 
is universal. This is because as $T_I\rightarrow\infty$, 
the initial thermal ensemble becomes a completely random ensemble. Consequently, 
the DE after a quench and the corresponding GE also become completely random 
ensembles and give identical results for all observables independently of the model
\cite{he_rigol_12_72,torres_santos_13}.

The question we address next is whether experimentally relevant observables, which are 
ensemble independent in thermal equilibrium in the thermodynamic limit, thermalize after 
a quench. Specifically, we consider the momentum distribution $m_k$ \cite{supmat} and 
the kinetic energy associated with nearest neighbor hoppings 
$K=-t\sum_i\langle\hat{b}^\dagger_i\hat{b}^{}_{i+1}\rangle$. 
In Fig.~\ref{fig:moment}(a), we show $m_k$ in the initial state ($T_I=2$), 
and in the DE and the corresponding GE after quenches with $t'=V'=0$ (integrable) 
and $t'=V'=0.5$ (nonintegrable). The DE and GE results are indistinguishable in 
the nonintegrable case, indicating thermalization, while they are 
clearly different at integrability, indicating the lack thereof.

When quantifying the differences between $m_k$ in the DE and the GE, we 
find that while the convergence of $m_k^\text{GE}$ is qualitatively similar to that of 
the observables analyzed previously, the same is not true for $m_k^\text{DE}$. As shown 
in the inset in Fig.~\ref{fig:moment}(a), for $T_I\gtrsim 2$, the results for 
$m_k^\text{GE}$ are converged within machine precision for all values of $t'=V'$ shown, 
while the ones for $m_k^\text{DE}$ are not. This indicates that clusters larger than 
those accessible here contribute to $m_k^\text{DE}$ in the thermodynamic limit and, as 
such, a careful scaling analysis is required to conclude whether thermalization occurs 
or not. We have found this to be true for other few-body observables such as $K$. 

In Figs.~\ref{fig:moment}(b) and \ref{fig:moment}(c), we show the relative 
difference $\delta(m)_l$ [defined in the same spirit as Eq.~\eqref{eq:d}, see 
Ref.~\cite{supmat}] between $m_k$ in the DE for order $l$ and in 
the last order in the GE vs $l$ for $T_I=2$ [Fig.~\ref{fig:moment}(b)]
and $T_I=10$ [Fig.~\ref{fig:moment}(c)]. The results are qualitatively 
similar at both temperatures but show different behavior 
depending on the value of $t'=V'$. At integrability, $\delta(m)_l$ approaches 
finite values as $l$ increases, so $m_k$ is not expected to thermalize in the 
thermodynamic limit. The convergence uncertainty in $m_k^\text{DE}$ 
is not significant in this case because $\delta(m)_{18}$ 
[Figs.~\ref{fig:moment}(b)] is almost 2 orders of magnitude greater than
$\Delta(S)_{17}$ [inset in Fig.~\ref{fig:moment}(a)].

\begin{figure}[!t]
    \includegraphics[width=0.465\textwidth]{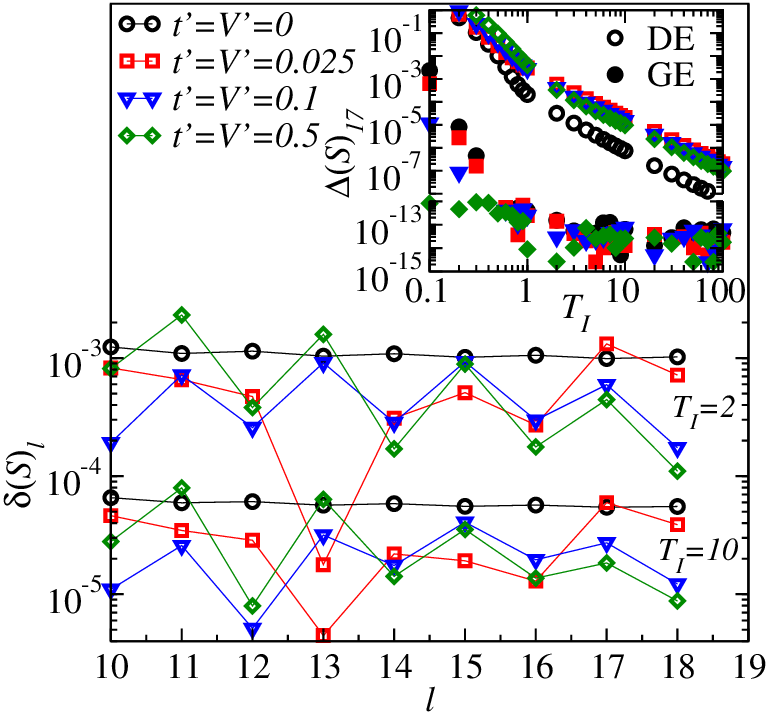}
\vspace{-0.1cm}
\caption{{\bf Entropy after a quench.} Relative difference $\delta(S)_l$
between $S^\text{DE}_l$ and $S^\text{GE}_{18}$ vs $l$ in quenches with $T_I=2$ 
(top four plots) and $T_I=10$ (bottom four plots). (Inset) Open (filled) symbols report the 
relative difference $\Delta(S)_{17}$ between $S^\text{DE}_l$ ($S^\text{GE}_l$) 
in the last 2 orders of the NLCE expansion vs $T_I$, for the same quenches as in the main
panel. The results in this figure are qualitatively similar to those in Fig.~\ref{fig:moment}.
For ease of display, there is a four decade gap in the $y$ axis of the inset.
}\label{fig:entropy}
\end{figure}

In the nonintegrable regime with $t'=V'>0.1$, we find that $\delta(m)_l$ in 
Figs.~\ref{fig:moment}(b) and \ref{fig:moment}(c) consistently decreases with 
increasing $l$ and that $\delta(m)_{18}\lesssim \Delta(m)_{17}$ 
[see the inset in Fig.~\ref{fig:moment}(a)]. These results suggest that 
nonzero values of $\delta(m)_l$ stem from the lack of convergence of $m_k^\text{DE}$ 
and will vanish as $l\rightarrow\infty$. Hence, our calculations provide strong evidence
that, in the thermodynamic limit, $m^\text{DE}_k=m^\text{GE}_k$.
When approaching the integrable point ($t'=V'=0.025$ in the plots), we find that 
the convergence of $m_k^\text{DE}$ worsens [see, e.g., $\Delta(m^\text{DE})_{17}$ 
in the inset in Fig.~\ref{fig:moment}(a)] and $\delta(m)_l$ vs $l$ exhibits erratic 
behavior~\cite{supmat}. For systems in 
equilibrium, such behavior is usually seen close to a phase transition, which suggests 
that in quantum quenches a phase transition to thermalization occurs as soon as one 
breaks integrability. This can be understood as, on approaching an integrable point, 
larger systems sizes are needed for the onset of eigenstate thermalization
\cite{santos_rigol_10_45,santos_rigol_10_49,neuenhahn_marquardt_12,steinigeweg_herbrych_13},
which results in larger cluster sizes needed for the series to converge to the thermal 
prediction. We have obtained qualitatively similar results when studying other few-body 
observables. In Fig.~\ref{fig:moment}(d) we plot results for $\delta(K)_l$ vs $l$, 
which are qualitatively similar to those for $\delta(m)_l$ in Figs.~\ref{fig:moment}(b) 
and \ref{fig:moment}(c).

Further evidence supporting the robustness of the picture above is provided
by the entropy. In the DE, the von Neumann entropy
$S^\text{DE}=-\sum_{\alpha} W_\alpha \ln (W_\alpha)$ has been argued to 
satisfy all properties expected of a thermodynamic entropy \cite{polkovnikov_11} 
and to agree (disagree) with the thermal entropy in quenches involving 
nonintegrable (integrable) systems where thermalization occurs (fails to occur) 
\cite{santos_polkovnikov_11_58}. As shown in Fig.~\ref{fig:entropy}, the 
relative differences $\delta(S)_l$ between $S^\text{DE}_l$ and $S^\text{GE}_{18}$ 
vs $l$ behave qualitatively similarly to $\delta(m)_l$ and $\delta(K)_l$ in 
Figs.~\ref{fig:moment}(b)--\ref{fig:moment}(d). The convergence of the NLCE 
for the entropy (inset in Fig.~\ref{fig:entropy}) is also qualitatively similar 
to that of $m_k$ (inset in Fig.~\ref{fig:moment}). Hence, our results provide 
further support the picture that $S^\text{DE}$ agrees (disagrees) with $S^\text{GE}$ 
when few-body observables thermalize (do not thermalize).

In summary, we have introduced NLCEs for the DE and shown that they can be used 
to study generic quenches in lattice systems in the thermodynamic limit. In the quenches 
studied here, NLCEs provided strong evidence that nonintegrable systems thermalize 
while integrable systems do not, and that a phase transition to thermalization may 
occur as soon as one breaks integrability. We plan to explore next 
whether NLCEs can be used to study dynamics, which would allow 
one to address fundamental questions related to prethermalization  
\cite{kinoshita_wenger_06,gring_kuhnert_12,berges_borsanyi_04,moeckel_kehrein_08,kollar_wolf_11} 
and to the time scales needed to observe thermalization in isolated systems.

\begin{acknowledgments}
This work was supported by the U.S. Office of Naval Research. We are grateful to D. Iyer, 
E. Khatami, L. F. Santos, and D. Weiss for comments on the manuscript.
\end{acknowledgments}

\newpage

\vspace*{0.3cm}

\onecolumngrid

\begin{center}

{\large \bf Supplementary Materials:
\\ Quantum Quenches in the Thermodynamic Limit}\\

\vspace{0.3cm}

Marcos Rigol\\
{\it Department of Physics, The Pennsylvania State University, University Park, Pennsylvania 16802, USA}\\

\end{center}

\vspace{0.6cm}

\twocolumngrid

\noindent{\it Hamiltonian.} The hard-core boson Hamiltonian reads
{\setlength\arraycolsep{0.5pt}
\begin{eqnarray}
\hat{H}&=&\sum_i \left\lbrace -t\left( \hat{b}^\dagger_i \hat{b}^{}_{i+1} + 
\textrm{H.c.} \right) 
+V\left( \hat{n}^{}_i-\dfrac{1}{2}\right)\left( \hat{n}^{}_{i+1}-\dfrac{1}{2}\right) 
\right.\nonumber\\
&-&\left.t'\left( \hat{b}^\dagger_i \hat{b}^{}_{i+2} + \textrm{H.c.} \right)  
+V'\left( \hat{n}^{}_i-\dfrac{1}{2}\right)\left( \hat{n}^{}_{i+2}-\dfrac{1}{2}\right)
\right\rbrace,\label{eq:hamil}
\end{eqnarray}}where $\hat{b}^\dag_i(\hat{b}^{}_i)$ denote the hard-core boson creation 
(annihilation) operators, and $\hat{n}^{}_i=\hat{b}^\dag_i\hat{b}^{}_i$ the number operator. 
In addition to the bosonic commutation relations $[\hat{b}^{}_i,\hat{b}^\dag _j]=\delta_{ij}$, 
those operators satisfy the constraints $\hat{b}^{\dag 2}_i=\hat{b}^2_i=0$, which prevent 
multiple occupancies of lattice sites in all physical states.\\

\noindent{\it Momentum distribution $m_k$.}
The momentum distribution function is defined as the Fourier transform  
$\hat{m}_k=(1/L)\sum_{jj'}e^{ik(j-j')}\hat{\rho}_{jj'}$ of the one-particle
density matrix $\hat{\rho}_{jj'}=\hat{b}_j^\dagger\hat{b}^{}_{j'}$. In our calculations, 
we compute $m_k$ in 100 equidistant $k$ points between $k=0$ and $\pi$. For $m_k$, in the 
same spirit of Eq.~(1) in the main text, we define
\begin{equation}
 \Delta (m^\text{ens})_l=\frac{\sum_k|(m_k)^\text{ens}_l-(m_k)^\text{ens}_{18}|}
 {\sum_k (m_k)^\text{ens}_{18}},
\end{equation}
where by ``ens'' we mean DE or GE, and, in the same spirit of Eq.~(2) in the main text, we define
\begin{equation}
 \delta (m)_l=\frac{\sum_k|(m_k)^\text{DE}_l-(m_k)^\text{GE}_{18}|}
 {\sum_k (m_k)^\text{GE}_{18}}.\label{eq:nkd}
\end{equation}

\noindent{\it Linked-cluster expansions.} In a linked-cluster expansion 
\cite{oitmaa_hamer_book_06}, the expectation value of an extensive observable 
$\hat{\mathcal{O}}$ per lattice site $\mathcal{O}=\langle\hat{\mathcal{O}}\rangle/L$ 
($L$ is the number of lattice sites), in the thermodynamic limit, is computed as the 
sum over the contributions from all clusters $c$ that can be embedded on the lattice
\begin{equation}
\label{eq:LCE1}
\mathcal{O}=\sum_{c}M(c)\times W_{\mathcal{O}}(c).
\end{equation}
$M(c)$ is the number of ways per site in which cluster $c$, with all sites connected, 
can be embedded on the lattice. [$M(c)$ is known as the multiplicity of $c$.] $W_{\mathcal{O}}(c)$ 
is the weight of that cluster for the observable $\mathcal{O}$, which is calculated using the 
inclusion-exclusion principle:
\begin{equation}
\label{eq:LCE2}
 W_{\mathcal{O}}(c)=\mathcal{O}(c)-\sum_{s \subset c} W_{\mathcal{O}}(s),
\end{equation}
where the sum runs over all connected sub-clusters of $c$ and
\begin{equation}
\label{eq:LCE3}
\mathcal{O}(c)={\textrm{Tr} [\hat{\mathcal{O}}\,\hat{\rho}_c]}/
{\textrm{Tr} [\hat{\rho}_c]}
\end{equation}
is the expectation value of $\hat{\mathcal{O}}$ calculated for the finite cluster $c$,
with many-body density matrix $\hat{\rho}_c$.\\

\begin{figure}[!t]
    \includegraphics[width=0.475\textwidth]{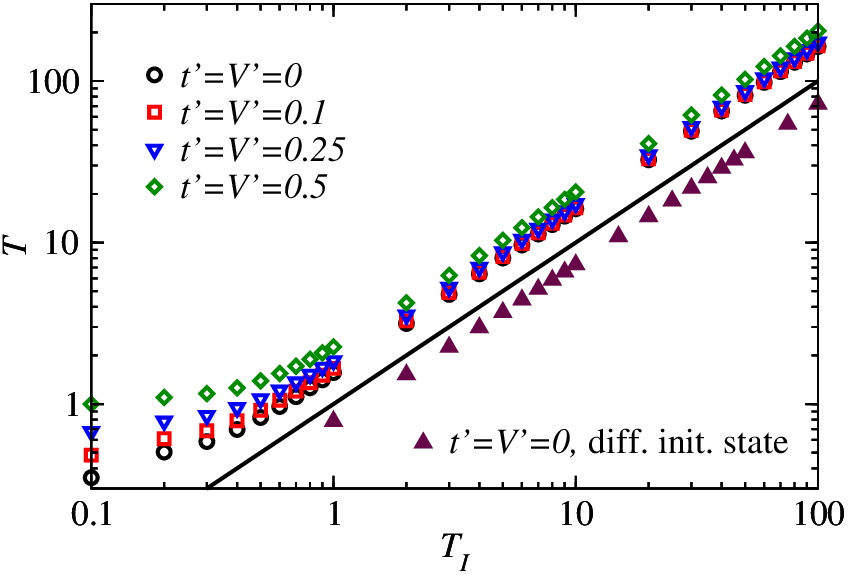}
\vspace{-0.1cm}
\caption{Effective temperature after the quench vs $T_I$, for the same quenches as in Fig.~1
in the main text. We also show results for a quench with with a different initial state with $t'=V'=0$ 
and $t_I=1.5$, $V_I=0.5$ ($t'_I=V'_I=0$ and $t=V=1$, as in all other quenches), which illustrates 
that $T$ can be lower than $T_I$. The straight line depicts $T=T_I$. 
}\label{fig:temp}
\end{figure}

\noindent{\it NLCE with maximally connected clusters.} 
An important feature of NLCEs, which is not present in other linked-cluster expansions,
is that one has quite some freedom in the selection of the building blocks used to carry 
out the expansion. One can use sites, bonds, and even squares or triangles depending on 
the geometry of the lattice \cite{rigol_bryant_06_25,rigol_bryant_07_30}. (A pedagogical 
introduction to NLCEs and their implementation can be found in Ref.~\cite{tang_khatami_13}.)
Here, we use the maximally connected clusters. For $l$ sites, the 
maximally connected cluster is the cluster with $l$ contiguous sites in which all nearest 
and next-nearest neighbor hoppings and interactions defined by the Hamiltonian are included. 
It is the only connected cluster with $l$ sites if $t'=V'=0$. Such an expansion is expected to be best 
suited when $t'=V'$ are small compared to $t$ and $V$. In our calculations, we carry out the 
NLCE computing observables in all maximally connected clusters with up to $l=18$.\\

\noindent{\it Effective temperature $T$ after the quench.} 
In Fig.~\ref{fig:temp}, we show $T$ for the quenches in Fig.~1 in the main text. While $T$ can be 
seen to be greater than $T_I$ in those quenches, this need not always occur. A quench, if $T_I>0$, 
can effectively cool a system. As an example, in Fig.~\ref{fig:temp} we also show $T$ for quenches 
in which $t'=V'=0$ after the quench, while $t_I=1.5$, $V_I=0.5$, where one can see that 
$T<T_I$ for $T_I\gtrsim1$.

\noindent{\it Dispersion of the density.} 
\begin{figure}[!t]
    \includegraphics[width=0.465\textwidth]{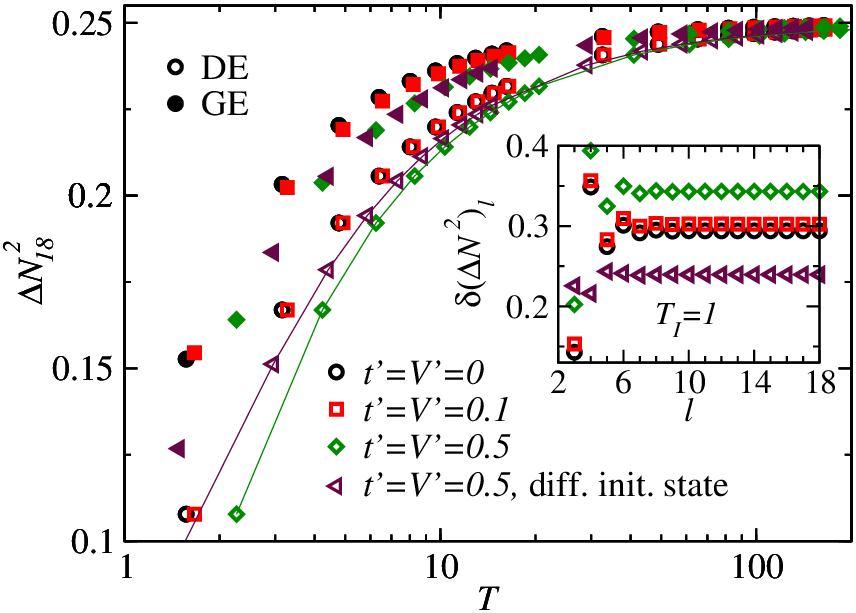}
\vspace{-0.1cm}
\caption{Last order ($l=18$) of the NLCE of $\Delta N^2$ in the 
DE (open symbols) and the GE (filled symbols) vs $T$, for quenches with 
$T_I\geq 1$. The quenches with a different initial state ($t'=V'=0.5$) 
have $t_I=0.5$, $V_I=1.5$, $t'_I=V'_I=0.5$, while $t=V=1$ as in the 
other quenches. Lines joining the data points for the quenches with $t'=V'=0.5$ 
are meant to guide the eye and show that $\Delta N^2$ in the DE depends on
the initial state. The inset shows $\delta(\Delta N^2)_l$ vs $l$ for 
$T_I=1$ for the same quenches as in the main panel.
}\label{fig:dfluctuations}
\end{figure}
In addition to the dispersion of the energy discussed in the main text, the dispersion
of the total number of particles (per site), $\Delta N^2=(\langle\hat{N}^2\rangle-\langle\hat{N}\rangle^2)/L$, 
also allow one to distinguish the DE from the GE. In thermal equilibrium, they depend 
on the ensemble used to compute them. $\Delta N^2=0$ in the microcanonical 
and canonical ensembles, while it can be different from zero only in the grand-canonical 
ensemble. $\Delta N^2$ is also of interest because, in the grand-canonical ensemble, the 
compressibility $\kappa=(1/L)\partial\langle\hat{N}\rangle/\partial\mu=\Delta N^2/T$.

In the main panel in Fig.~\ref{fig:dfluctuations}, we plot $\Delta N^2$
in the DE (empty symbols) and in the GE (filled symbols) vs $T$ for quenches
with $T_I\geq 1$. These results, particularly the ones at the lowest temperatures, 
make it apparent that $\Delta N^2$ is different in the DE and the GE.
Moreover, for quenches with different initial states but the same final Hamiltonian 
(with $t'=V'=0.5$), $\Delta N^2$ in the DE can be seen to depend on the initial 
state. The relative differences $\delta(\Delta N^2)_l$ between $\Delta N^2$ in the DE 
for order $l$ and in the last order in the GE are plotted vs $l$ in the inset in 
Fig.~\ref{fig:dfluctuations}. They show that the nonzero differences seen 
in the main panel between $\Delta N^2$ in the DE and the GE are fully 
converged and are thus expected to be the ones in the thermodynamic limit.

\begin{figure}[!t]
    \includegraphics[width=0.47\textwidth]{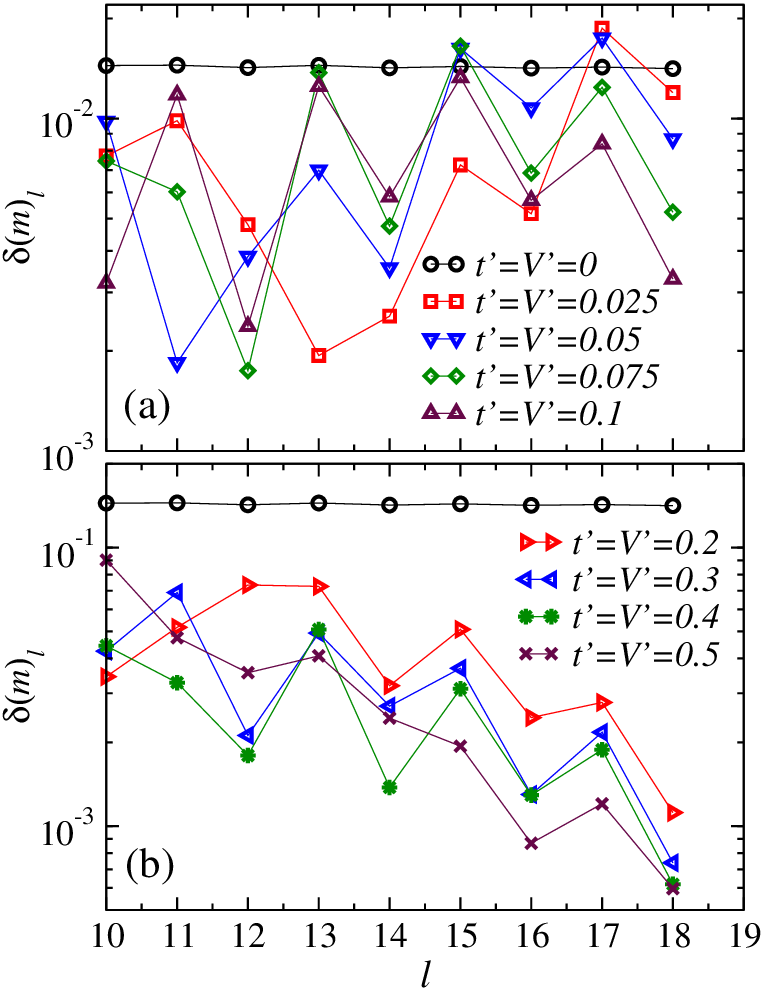}
\vspace{-0.1cm}
\caption{Relative differences between $(m_k)^\text{DE}_l$ and $(m_k)^\text{GE}_{18}$
vs $l$ in nine sets of quenches with $T_I=2$.
}\label{fig:momconv}
\end{figure}

\noindent{\it Convergence of the DE results close to the integrable point.}
In the main text we discussed that as one approaches the integrable point $\delta(m)_l$ 
vs $l$ exhibits erratic behavior, and that this is the result of the worsening of the 
convergence of $m_k^\text{DE}$. In systems in equilibrium such a behavior is usually seen 
close to phase transitions, so we argued that a phase transition to thermalization may occur 
as soon as one breaks integrability. 

In order to supplement the results in Fig.~3 in the main text, so that such an erratic 
behavior on approaching integrability is better seen by comparing to results as 
one departs from integrability, in Fig.~\ref{fig:momconv} we plot $\delta(m)_l$ vs $l$ for 
almost three times as many values of $t'=V'\neq0$ as those in the main text. In 
Fig.~\ref{fig:momconv}(a) one can see that, for $t'=V'=0.025$ and $t'=V'=0.05$, 
$\delta(m)_l$ first decreases and then increases as $l$ increases. For $t'=V'=0.075$, 
again $\delta(m)_l$ first decreases, then increases, and for the largest values of $l$ 
it appears to decrease again. The latter behavior becomes more evident for $t'=V'=0.1$. 
All results for nonzero $t'=V'\lesssim1$ are also characterized by very large 
oscillations in the values of $\delta(m)_l$, which are not present at integrability. 
On the other hand, the results for $t'=V'>1$ in Fig.~\ref{fig:momconv}(b) offer a different 
picture. $\delta(m)_l$ consistently decreases with increasing $l$, which supports the view 
that thermalization occurs.

\begin{figure*}[!t]
    \includegraphics[width=0.8\textwidth]{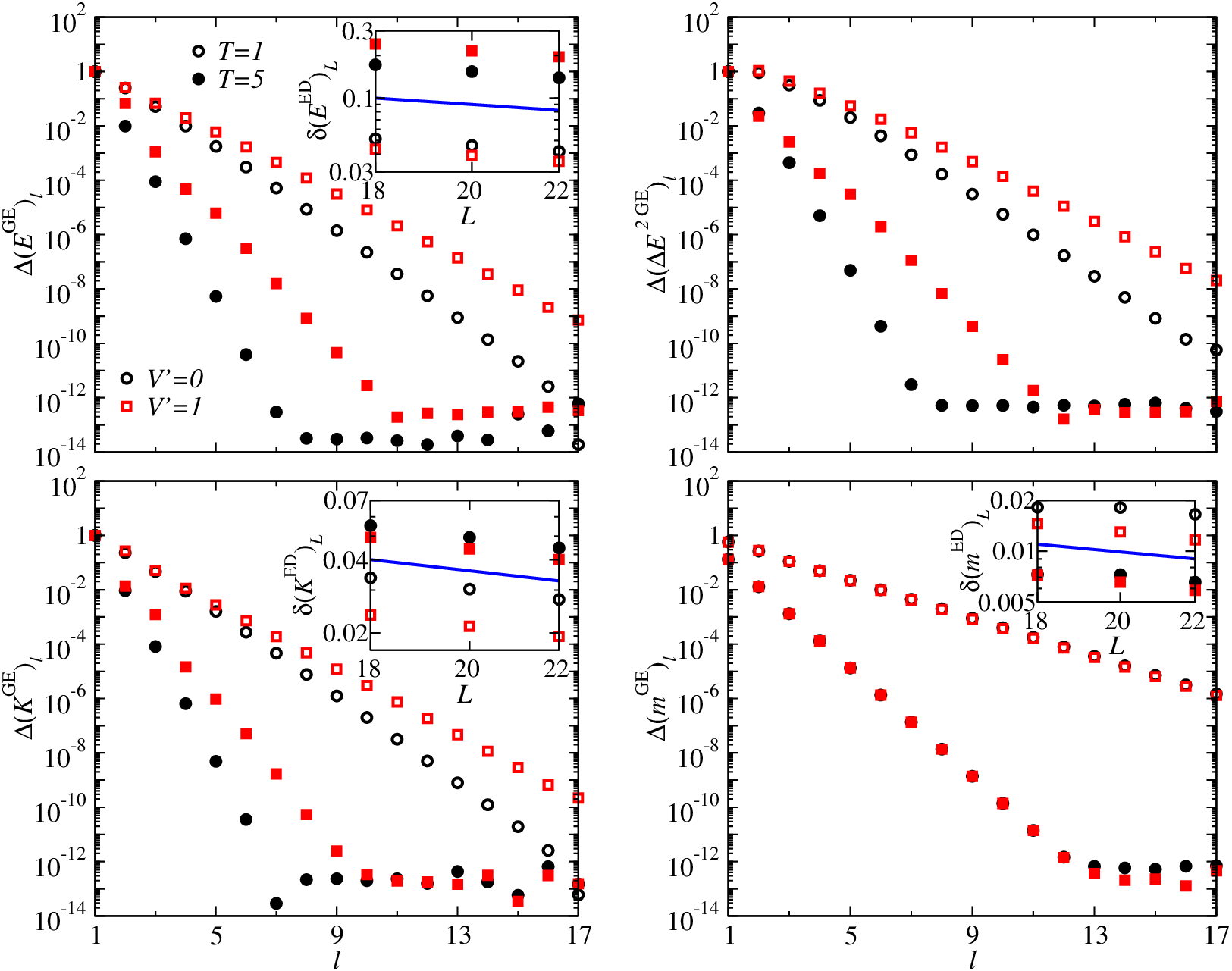}
\vspace{-0.1cm}
\caption{Open (filled) symbols report the relative differences $\Delta(E^\text{GE})_l$ (a), 
$\Delta(\Delta E^{2\,\text{GE}})_{l}$ (b), $\Delta(K^\text{GE})_{l}$ (c), and 
$\Delta(m^\text{GE})_{17}$ (d) vs $l$ for $T=1$ ($T=5$). Results are presented for 
$V'=0$ (circles) and $V'=1.0$ (squares). In all cases one can see an exponential 
decrease of the differences with increasing cluster size. In the insets in (a), (c), and (d) 
we show log-log plots of relative differences between results of full exact diagonalization 
in systems with periodic boundary conditions ($L=18,\,20$ and 22 sites) and NLCE results 
with $l=18$. In this case one can see that, as expected, the results are consistent 
with power law scaling in $L$. We have included straight lines in the insets explicitly 
depicting $1/L$ scaling.
}\label{fig:conv}
\end{figure*}

\noindent{\it Convergence of NLCEs vs exact diagonalization.} 
In Fig.~\ref{fig:conv}, we show relative differences between NLCE results for four observables 
when all contributions from clusters with up to $l$ sites are added and the results for $l=18$ 
(the highest order in the NLCE calculation that we have computed) vs $l$. Those relative differences 
were defined in Eq.~(1) in the main text and in Eq.~\eqref{eq:nkd} here. We took as Hamiltonian 
Eq.~\eqref{eq:hamil} when $t'=0$, which was systematically studied using exact diagonalization 
in Ref.~\cite{santos_rigol_10_49}. One can see in all panels in Fig.~\ref{fig:conv} that the 
relative differences decrease exponentially fast with the order $l$ of the NLCE calculation
(note that results are presented for two temperatures $T=1$ and $T=5$).

In the insets in Figs.~\ref{fig:conv}(a), \ref{fig:conv}(c), and \ref{fig:conv}(d), 
we report relative differences between results obtained using full exact diagonalization (ED)
in systems with periodic boundary conditions ($L=18,\,20$, and 22 sites) \cite{santos_rigol_10_49} 
and the NLCE results for $l=18$
\begin{equation}
 \delta (\mathcal{O}^\text{ED})_L=\frac{|\mathcal{O}^\text{ED}_L-\mathcal{O}^\text{GE}_{18}|}
 {|\mathcal{O}^\text{GE}_{18}|}.\label{eq:ed}
\end{equation}
and 
\begin{equation}
 \delta (m^\text{ED})_L=\frac{\sum_k|(m_k)^\text{ED}_L-(m_k)^\text{GE}_{18}|}
 {\sum_k (m_k)^\text{GE}_{18}},\label{eq:nked}
\end{equation}
where the sum in Eq.~\eqref{eq:nked} is restricted to the values of $k$ that are available 
in the specific cluster with periodic boundary conditions used in the exact diagonalization
calculation. These differences exhibit a scaling that is consistent with $1/L$, as expected. 
They make evident that the scaling (and ultimately the accuracy) of the results obtained 
using NLCEs and ED are fundamentally different, namely, exponential (NLCEs) vs power law (ED).

\end{document}